%Paper: cond-mat/9411073
%From: bhaduri@smiley.physics.mcmaster.ca (Dr. R. K. Bhaduri)
%Date: Thu, 17 Nov 1994 15:25:42 -0500

\magnification=1250
\overfullrule=0pt
\baselineskip=19pt

\def\th{thermodynamic }
\def\po{potential }

\centerline{\bf Thomas-Fermi Method For Particles Obeying Generalized}
\centerline{\bf Exclusion Statistics}
\vskip.5in

\centerline{Diptiman Sen}
\vskip .1in

\centerline{\it Centre for Theoretical Studies, Indian Institute of
Science,}
\centerline{\it Bangalore 560012, India}
\vskip.2in

\centerline{R. K. Bhaduri}
\vskip .1in

\centerline{\it Department of Physics and Astronomy, McMaster University,}
\centerline{\it Hamilton, Ontario L8S 4M1, Canada}
\vskip.3in

\line{\bf Abstract \hfill}
\vskip .1in

We use the Thomas-Fermi method to examine the thermodynamics of
particles obeying Haldane exclusion statistics. Specifically, we study
Calogero-Sutherland particles placed in a given external potential in one
dimension. For the case of a simple harmonic potential (constant density
of states), we obtain the exact one-particle spatial
density and a {\it closed} form for the equation of state at finite
temperature, which are both new results. We then solve the problem of
particles in a $x^{2/3} ~$ potential (linear density of states) and show
that  Bose-Einstein condensation does not occur for any statistics other
than bosons.

\vskip .4in

\line{PACS numbers: ~5.30.-d, ~71.45.Jp \hfill}

\vfill
\eject

A recent definition of generalized exclusion statistics by Haldane has
aroused considerable interest [1]. The definition is based on the rate
at which the number of available states in a system of fixed size
decreases as more and more particles are added to it. We introduce a
statistics parameter $\alpha$ and assign the values $\alpha=0$ and $1$
to bosons and fermions respectively, because the addition of one
particle reduces the number of available states by $\alpha$.
Some examples of particles obeying such generalized statistics are
spinons in an antiferromagnetic spin chain with inverse square exchange
[1,2] and two-dimensional anyons residing in the lowest Landau level
in a strong magnetic field [3-5].

More recently, the statistical distribution function for such systems
has been obtained [5,6]. If $\mu$ is the chemical \po ,
$\epsilon$ is the energy of a single-particle state and $T$ is the
temperature, then $n(\epsilon, \mu, T)$ satisfies
$$(n^{-1} ~-~ \alpha)^{\alpha} ~(n^{-1} ~+~ 1~-~ \alpha)^{1- \alpha} ~
=~ \exp [(\epsilon ~-~ \mu)/T ~].
\eqno(1)$$
(We set Planck's and Boltzmann's constants equal to one). Let
$g(\epsilon)$ be the density of states so that the total number of
particles and the energy are given by
$$\eqalign{N ~&= ~\int ~ d \epsilon ~g(\epsilon) ~n(\epsilon) \cr
{\rm and} \quad E ~&= ~\int ~ d \epsilon ~g(\epsilon) ~n(\epsilon) ~
\epsilon. \cr}
\eqno(2)$$
Using Eq. (1), the \th \po $\Omega (\mu, T)$ satisfying
$({\partial \Omega} / {\partial \mu})_T ~=~ -~N$ can be shown to be
given by
$$\Omega ~=~ -~T ~\int ~ d\epsilon ~ g(\epsilon) ~\log [1 ~+~ \exp
(-~ {\tilde \epsilon} /T) ~],
\eqno(3)$$
where ${\tilde \epsilon} (\epsilon, \mu, T)$ is defined by the relation
$${\tilde \epsilon} ~+~ T ~(1 ~-~ \alpha) ~\log [1 ~+~ \exp (-~
{\tilde \epsilon} /T)~] ~=~ \epsilon ~-~ \mu.
\eqno(4)$$
In terms of $\tilde \epsilon$, $n$ is given by
$$n ~=~ {1 \over {\exp ({\tilde \epsilon} /T)} ~+~ \alpha}.
\eqno(5)$$

If we keep $T$ fixed and let $\mu \rightarrow ~-~ \infty$, we find that
${\tilde \epsilon} \rightarrow \epsilon - \mu$ and $n(\epsilon)
\rightarrow ~ \exp [(\mu - \epsilon)/T] ~$ independent of $\alpha$. In this
limit, therefore, $N, ~ \Omega$ and $E$ all go to zero as $\exp (\mu /T)$.
We now use the identity
$${{\partial n} \over {\partial \epsilon}} ~+~ {{\partial n} \over
{\partial \mu}} ~=~ 0,
\eqno(6)$$
which follows from Eq. (1) if T is held fixed. Now suppose that $g
(\epsilon) ~\sim ~ {\epsilon}^{a} ~$. Inserting (6) in (2), we obtain
$$\biggl({{\partial E} \over {\partial \mu}} \biggr)_T ~=~ (a ~+~ 1)~ N.
\eqno(7)$$
Hence $E+ (a+1) \Omega$ is independent of $\mu$. Since $E+(a+1)
\Omega$ is zero at $\mu = - \infty$, it must be zero
for all $\mu$ and $T$. This relation will be used below.

The simplest example of particles obeying the generalized exclusion
statistics is provided by the Calogero-Sutherland (CS) model [7,8].
This one-dimensional model has the Hamiltonian
$$H ~=~ {1 \over {2m}} ~\sum_i ~ p_i^2 ~+~ {{\alpha (\alpha -1)} \over
m} ~\sum_{i<j} ~{1 \over {(x_i ~-~ x_j )^2 }}.
\eqno(8)$$
Here $\alpha \ge 0$ with the understanding that the wave function
vanishes as $\vert x_i ~-~ x_j ~\vert^{\alpha} ~$ whenever the
particles i and j approach each other. Note that $\alpha=0$ and $1$
yield free particles (bosons and fermions repectively) while the
maximally attractive \po occurs for $\alpha=1/2 ~$ (semions).
Using the \th Bethe ansatz [9] and other means,
it has been argued that the CS system
can be thought of as an ideal gas in the sense that the particles only
have statistical interactions amongst each other [6,10,11].

While the CS model has been studied in great detail over the years, we
would like to examine it here from the view point of exclusion
statistics. The CS model has so far been exactly solved only in an
external simple harmonic \po $V(x) ~=~ m \omega^2 x^2 /2 ~$.
However, once we realise that the
particles only undergo statistical interactions amongst each other, we can
develop other techniques to study the \th of the CS system placed in
{\it any} external \po . One such technique is the Thomas-Fermi (TF)
method which has been used to study fermionic systems for a long time [12].
In addition to being exact in the large-$N$ limit, the TF method does not
require a knowledge of the energy spectrum and wave function of the
system. Here we modify the method by using the distribution in
(1) rather than the Fermi-Dirac distribution while continuing to use
the usual phase space description of quantum systems.
We show below that this procedure leads to the well-known expressions
for the one-particle spatial density and the ground state
energy for the many-body system in a simple harmonic \po
at zero temperature [7,8].
Encouraged by this, we analyse the situation at finite temperaure and
obtain the one-particle density which is a new result. We also find
closed forms for the \th quantities $\mu$ and
$\Omega$, in contrast to previous expressions in the literature which
are given as high-temperature series. Finally, to demonstrate the power
of the TF method, we study the CS model placed in a $x^{2/3} ~$
\po which has $g(\epsilon) ~\sim ~\epsilon$. Once again, we
derive the \th quantities for all $\alpha$ and $T$. Since the
$g(\epsilon)$ is the same as for free particles in four dimensions, it
is natural ask whether Bose-Einstein (BE) condensation occurs at low
temperature for a {\it finite} range of $\alpha$. It is shown that
condensation occurs {\it only} for bosons ($\alpha =0$) and not for any
positive $\alpha$.

The TF density of states is obtained from the integral
$$\eqalign{g(\epsilon) ~&=~ \int {{dx dp} \over {2 \pi}} ~\delta
(\epsilon ~-~ {{p^2} \over {2m}} ~-~ V(x)) \cr
&=~ \int ~{dx \over \pi} ~{{\theta (\epsilon ~-~ V(x))} \over {\sqrt
{2(\epsilon ~-~ V(x))/m}}} ,\cr}
\eqno(9)$$
where $\theta (y) = 1$ if $y>0$ and $0$ if $y<0$.
The TF expression for the one-particle density is
$$\rho (x,T) ~=~ \int ~ {dp \over {2 \pi}} ~ n(\epsilon,T),
\eqno(10)$$
where $\epsilon=p^2 /2m ~+~ V(x)~$ and $~\int ~dx ~\rho (x,T) = N$.

At $T=0$, Eq. (1) implies that there is a Fermi energy $\mu (0)$ such that
$$\eqalign{n(\epsilon) ~&=~ 1  / \alpha \quad {\rm if} \quad
\epsilon ~<~ \mu (0), \quad {\rm and} \cr
&= ~0 \quad {\rm if} \quad \epsilon ~>~ \mu (0). \cr}
\eqno(11)$$
Thus $\alpha < 1$ (or $> 1$) corresponds to less (or more) exclusion
than fermions. We now study the harmonic \po problem. From (9),
$g(\epsilon)= 1/ \omega ~$. Therefore the
Fermi and ground state energies are given by
$$\eqalign{\mu (0) ~&=~ \alpha \omega N \cr
{\rm and} \quad E (0) ~&=~ {1 \over 2} ~N \mu (0 ). \cr}
\eqno(12)$$
Eqs. (10, 11) yield the well-known semicircle law [8,13]
$$\rho (x,0) ~=~ {1 \over {\pi \alpha}} ~(2m \mu (0) ~-~ m^2 \omega^2
x^2 ~)^{1/2}
\eqno(13)$$
for $x^2 ~\le ~x_o^2 ~=~ 2 \mu (0) /m \omega^2 ~$. The \th limit for
this system is obtained by taking $N \rightarrow \infty$ holding $\mu
(0)$ and therefore $\omega N$ fixed. (This is analogous to keeping the
density in a box fixed with $1 / \omega$ playing the role of the volume).

At finite temperature, we find from Eqs. (2, 6) that
$$\omega ~\biggl({{\partial N} \over {\partial \mu}} \biggr)_T ~=~ n(0),
\eqno(14)$$
where $n(0)$ is the solution of (1) with $\epsilon=0$. This has the
solution
$$\mu (N,T) ~=~ \alpha \omega N ~+~ T ~\log [1- \exp (- \omega N/T)].
\eqno(15)$$
(Here we used the boundary condition that for fixed $T$, $N \rightarrow
0$ as $\mu \rightarrow - \infty$). Now
$$\biggl({{\partial \Omega} \over {\partial N}} \biggr)_T ~=~ - N ~
\biggl({{\partial \mu} \over {\partial N}} \biggr)_T
\eqno(16)$$
implies that [14]
$$-~ \Omega ~=~ {1 \over 2} \alpha \omega N^2 ~+~ {T^2 \over
\omega} ~\int_0^{\omega N /T} ~ dy ~{y \over {e^y ~-~ 1}}.
\eqno(17)$$
Since $E=- \Omega$, the entropy $S$ (given by $\Omega = E - T S - \mu
N ~$) is independent of $\alpha$. Eq. (17) has a high-temperature
expansion in $\omega N /T ~$ of the form
$$- ~\Omega ~=~ NT ~+~ {1 \over 2} ~\omega N^2 ~(\alpha ~-~ {1
\over 2}) ~+~ ... ~.
\eqno(18)$$
This resembles the virial expansion for the equation of state in a
box and it has the remarkable property that only the second term
depends on $\alpha$ [11]. However, this expansion only has a
finite radius of convergence ($2 \pi$). Eq. (17) has an alternative
expansion in powers of $\exp (- N \omega /T ~)$ which is valid for
{\it all} $T$. For instance, we may read off the low-temperature result
$$-~ \Omega ~=~ {1 \over 2} \alpha \omega N^2 ~+~ {\pi^2
\over 6} {T^2 \over \omega}
\eqno(19)$$
plus exponentially small terms.

We may now obtain the finite temperature density $\rho (x,T)$ from Eqs.
(1) and (10). This is somewhat difficult to compute explicitly except for
special values of $\alpha$. For instance,
$$\eqalign{n(\epsilon) ~&=~ {2 \over {\sqrt {1+4 \gamma^2}}} \quad
{\rm for} \quad \alpha ~=~ 1/2 \quad {\rm and} \cr
{}~&=~ {1 \over 2} ~-~ {1 \over 2} {\sqrt {\gamma
\over {4 + \gamma}}} \quad {\rm for} \quad \alpha ~=~ 2, \cr}
\eqno(20)$$
where $\gamma = \exp [(\epsilon - \mu)/T] ~$. In Fig. 1,
we show $\rho (x,T)$ at $T=\omega N/2$ for $\alpha = 1/2, ~1$ and $2$.
For convenience, we have rescaled $x \rightarrow x/ x_o ~$ and $\rho
\rightarrow ~{\rho (x,T)} / {\rho (0,0)}$ so that
$$\eqalign{\rho (x,0) ~&=~ {\sqrt {1- x^2 }} ~\theta (1- x^2) \cr
{\rm and} \quad \int_0^{\infty} ~dx ~\rho (x,T) ~&=~ \pi / 4 \cr}
\eqno(21)$$
for all $\alpha$.
Although $T$ is fixed in Fig. 1, its ratio to $\mu (0)= \alpha \omega N$
is greater for smaller $\alpha$. This explains why the curve for $\alpha
= 1/2$ is spread out to the largest extent. We have also exhibited the
$T=0$ (or $\alpha \rightarrow \infty$) semicircle for comparison.

We now apply the TF idea to a CS system placed in a different \po
which has not been studied before. An interesting and exactly solvable
TF case is the \po
$$V(x) ~=~ {1 \over 2} ~\Bigl(9 m \eta^2 x^2 \Bigr)^{1/3}.
\eqno(22)$$
The TF density of states is given by $g(\epsilon) ~=~ \epsilon /
\eta ~$. Hence $E=~-2 \Omega$. At $T=0$, we find that
$$\eqalign{\mu (0) ~&=~ \Bigl(2 \alpha \eta N \Bigr)^{1/2} \cr
{\rm and} \quad E (0) ~&=~ {2 \over 3} ~N \mu (0). \cr}
\eqno(23)$$
The \th limit is again defined by taking $N \rightarrow \infty$ keeping
$\eta N$ fixed. To study finite temperature, we define an auxiliary
variable $$\nu ~=~ \int_0^{\infty} ~d \epsilon ~n (\epsilon).
\eqno(24)$$
Eq. (6) implies that
$$\eqalign{\eta ~\biggl({{\partial N} \over {\partial \mu}}\biggr)_T ~&=~
\nu \cr
{\rm and} \quad \biggl({{\partial \nu} \over {\partial \mu}}\biggr)_T ~&=~
n(0). \cr}
\eqno(25)$$
As in Eqs. (15, 17), we deduce that
$$\eqalign{\mu (\nu, T) ~&=~ \alpha \nu ~+~ T ~\log [1~-~ \exp(- \nu /T)] \cr
{\rm and} \quad \eta N(\nu, T) ~&=~ {1 \over 2} ~ \alpha \nu^2 ~+~
T^2 ~\int_o^{\nu /T} ~ dy~ {y \over {e^y ~-~ 1}} .\cr}
\eqno(26)$$
Using (7), we can compute the energy
$$E(\nu, T) ~=~ 2 ~\int_0^{\nu} ~dy~ N(y,T) ~{\partial \over {\partial y}} ~
\mu (y,T).
\eqno(27)$$
Thus $\mu, ~N$ and $E$ are all known in terms of $\nu$ and $T$.

Since a particle moving in a two-dimensional harmonic oscillator \po
or moving freely in four dimensions also has a linear density of states,
we may ask whether our system exhibits BE condensation at low
temperature for a {\it finite} range of $\alpha$. If $\alpha =0$, we see that
(26) has no solution once $T$ falls below
$$T_c ~=~ {\sqrt {6 \eta N}} / \pi .
\eqno(28)$$
For $T \le T_c ~$, $\nu$ and $\mu$ stay at $\infty$ and $0$ respectively.
Therefore the second equation in (26) can only be satisfied
if there is a macroscopic occupation $N (0)$ of the $\epsilon =0$ state,
that is, if $\eta N ~=~ \eta N(0) ~+~ \pi^2 T^2 /6 ~$.

However, for any $\alpha > 0$, Eq. (26) has a solution right down to $T=0$
and there is no condensation at finite temperature. In Fig. 2, we
present $\mu$ vs $T$ with
$\eta N=1$ and $\alpha =0, ~0.02, ~0.1$ and $0.25$. It will be seen that $\mu
(T)$ is nonanalytic only for $\alpha =0$ at $T= {\sqrt 6} / \pi =
0.7797 ... ~$.

To summarize, we have developed a TF method to study the thermodynamics of
particles obeying the generalized statistics of Eq. (1) in any given external
\po . We can numerically compute the one-particle density for any
$\alpha$ and $T$. If $g(\epsilon)$ varies as a (nonnegative) integer power
of $\epsilon$, we can solve the thermodynamics exactly. For a linear
density of states, there is no BE condensation if $\alpha > 0$. This
also implies that a $1/x^2 ~$ interaction between bosons, no matter
how weak, destroys BE condensation.

One of us (D. S.) thanks the Physics and Astronomy Department of McMaster
University for its hospitality during the course of this work. This
research was supported by a grant from NSERC (Canada).

\vfill
\eject

\line{\bf References \hfill}
\vskip .2in

\noindent
\item{1.}{F. D. M. Haldane, Phys. Rev. Lett. {\bf 67} (1991) 937.}

\noindent
\item{2.}{F. D. M. Haldane, Phys. Rev. Lett. {\bf 60} (1988) 635;
B. S. Shastry Phys. Rev. Lett. {\bf 60} (1988) 639.}

\noindent
\item{3.}{M. V. N. Murthy and R. Shankar, Phys. Rev. Lett. {\bf 72}
(1994) 3629.}

\noindent
\item{4.}{A. Dasni\`eres de Veigy and S. Ouvry, Phys. Rev. Lett. {\bf
72} (1994) 600.}

\noindent
\item{5.}{Y.-S. Wu, Phys. Rev. Lett. {\bf 73} (1994) 922.}

\noindent
\item{6.}{S. B. Isakov, Phys. Rev. Lett. {\bf 73} (1994) 2150.}

\noindent
\item{7.}{F. Calogero, J. Math. Phys. {\bf 10} (1969) 2191, 2197.}

\noindent
\item{8.}{B. Sutherland, J. Math. Phys. {\bf 12} (1971) 246, 251;
Phys. Rev. A {\bf 4} (1971) 2019.}

\noindent
\item{9.}{C. N. Yang and C. P. Yang, J. Math. Phys. {\bf 10} (1969)
1115.}

\noindent
\item{10.}{D. Bernard and Y.-S. Wu, preprint UU-HEP/94-03
(unpublished).}

\noindent
\item{11.}{M. V. N. Murthy and R. Shankar, preprint IMSc-94/24
(unpublished).}

\noindent
\item{12.}{N. H. March, {\it Self-Consistent Fields in Atoms}
(Pergamon Press, New York, 1975).}

\noindent
\item{13.}{M. L. Mehta, {\it Random Matrices} (Academic Press,
New York, 1991).}

\noindent
\item{14.}{D. Sen, Nucl. Phys. B {\bf 360} (1991) 397.}

\vfill
\eject

\line{\bf Figure Captions \hfill}
\vskip .2in

\noindent
\item{1.}{The one-particle density $\rho (x,T)$ vs $x$ in a harmonic
oscillator \po with $T = \omega N /2$ and $\alpha= 1/2, ~1$ and
$2$. The $T=0$ (or $\alpha \rightarrow \infty$ ) semicircle is shown
as a solid line.}

\noindent
\item{2.}{The chemical \po $\mu$ vs $T$ for a linear density
of states with $\eta N=1$ and $\alpha=0, ~0.02, ~0.1$ and $0.25$.
The solid line ($\alpha=0$) is nonanalytic at $T= {\sqrt 6} / \pi $.}

\end